# Bilinear magneto-electric resistance as a probe of three-dimensional spin texture in topological surface states

Pan He[1†], Steven S.-L. Zhang[2†], Dapeng Zhu[1], Yang Liu[1], Yi Wang[1], Jiawei Yu[1], Giovanni Vignale[2] and Hyunsoo Yang[1]*

[1]*Department of Electrical and Computer Engineering, and NUSNNI, National University of Singapore, 117576, Singapore*

[2]*Department of Physics and Astronomy, University of Missouri, Columbia Missouri 65211, USA*

[†]These authors contributed equally to this work. *e-mail: <u>eleyang@nus.edu.sg</u>

**Surface states of three-dimensional topological insulators exhibit the phenomenon of spin-momentum locking, whereby the orientation of an electron spin is determined by its momentum. Probing the spin texture of these states is of critical importance for the realization of topological insulator devices, however the main technique available so far is the spin- and angle-resolved photoemission spectroscopy. Here we reveal a close link between the spin texture and a new kind of magneto-resistance, which depends on the relative orientation of the current with respect to the magnetic field as well as the crystallographic axes, and scales linearly with both the applied electric and magnetic fields. This *bilinear magneto-electric resistance* can be used to map the spin texture of topological surface states by simple transport measurements. For a prototypical $Bi_2Se_3$ single layer, we can map both the in-plane and the out-of-plane components of the spin texture – the latter arising from hexagonal warping.**

**Theoretical calculations suggest that the bilinear magneto-electric resistance originates from the conversion of a non-equilibrium spin current into a charge current under the application of the external magnetic field.**

The discovery of three-dimensional (3D) topological insulators (TIs) has triggered an outburst of research activities aimed at understanding the physical properties of this new state of matter[1-4], and exploring its applications to optoelectronics and spintronics[5, 6]. The most remarkable property of TIs is their topologically protected surfaces states characterized by a tight correlation between spin orientation and momentum, known as spin-momentum locking[1-4]. The resulting spin texture has been mapped by spin- and angle-resolved photoemission spectroscopy (SR-ARPES)[7-9]. When the topological surface states (TSS) are hexagonally deformed (Fig. 1a)[10, 11], the spin texture acquires a momentum-dependent out-of-plane component[10]. This hexagonally warped helical spin texture, depicted in Fig. 1b, has been observed in various TI materials[12-16] by SR-ARPES. In addition to the optical methods[13-17], electrical detection of the spin-polarized TSS has been achieved with the use of ferromagnetic contacts as spin detectors[18-22]. However, only the in-plane spin component could be detected by this method[18-22].

The magneto-transport properties of 3D TIs have also been extensively studied[23-29]. Various types of magneto-resistance in non-magnetic 3D TIs have been identified including the weak (anti)localization induced magneto-resistance[23, 24], the non-saturating linear magneto-resistance[25-27], and the anisotropic magneto-resistance[28, 29], which all appear in the linear response of the current to the applied electric field. Recently, a large unidirectional magneto-resistance (UMR) was measured in heterostructures consisting of non-magnetic and magnetic topological insulators[30], which exhibits a linear dependence on the current density, however, the presence of a magnetic layer was deemed to be essential for this nonlinear magneto-resistance. So far, no correlation between the magneto-resistance and the spin texture of the TSS has been established, even though a tight link between the charge and spin currents is expected due to spin-momentum locking.

Here, we report the observation, in a single layer of nonmagnetic 3D TI $Bi_2Se_3$, of a new magneto-resistance effect which scales linearly with the applied electric current and magnetic field: we thus name it "bilinear magneto-electric resistance". We show that a mapping between momentum-dependent spin textures and the angular dependence of the bilinear magneto-electric resistances can be established, which enables transport measurements of the 3D spin texture in the TSS with hexagonal warping.

**Observation of the bilinear magneto-electric resistance**

The samples studied in this work are 20 quintuple layer (QL) $Bi_2Se_3$ films, which are capped by MgO (2 nm)/$Al_2O_3$ (3 nm) and patterned into Hall bars of length $l = 100$ µm and width $w = 20$ µm. The current channels ($x$ axis) of the Hall bar devices were designed along different crystal directions on the same films. To measure the current-independent first harmonic $R_\omega$ and the current-dependent second harmonic $R_{2\omega}$ of the longitudinal resistance[31] (see Supplementary Information S1), a.c. currents were injected into the devices, and the first ($V_\omega$) and second ($V_{2\omega}$) harmonic longitudinal voltages ($V_{xx}$) were recorded simultaneously. The measurements were performed while rotating the applied magnetic field $H$ in the $xy$, $zy$, and $zx$ planes as schematically shown in Fig. 2a-c, respectively.

In Figs. 2d-f, we show the angular dependences of $R_\omega$ (blue) and $R_{2\omega}$ (red) when a fixed magnetic field of 9 T is swept in the *xy*, *zy*, and *zx* planes, respectively. For scans in all three planes, the linear longitudinal resistance $R_\omega$ is anisotropic[28, 29], and exhibits a sinusoidal angular dependence with a period of 180°. This angular dependence shows that $R_\omega$ is invariant under the reversal of the magnetic field.

More interestingly, we observe a novel anisotropic behavior in the nonlinear resistance $R_{2\omega}$. For the three scans shown in Fig. 2d-f, $R_{2\omega}$ exhibits sinusoidal angular dependences with a period of 360°, but with different phases in different planes. Contrary to $R_\omega$, $R_{2\omega}$ changes its sign when reversing either the current or the magnetic field polarity (see Supplementary Information S3). The measured $R_{2\omega}$ indicates the existence of a current-dependent unidirectional resistance in $Bi_2Se_3$, which has also been confirmed by dc measurements[30, 32, 33] (see Supplementary Information S4,5). We further investigate $R_{2\omega}$ by varying the direction of the current with respect to the $\overline{\Gamma K}$-axis of TI. We show that there exists a mapping between the spin canting angle at a given ***k***-point on the Fermi contour of the TSS and the magnetic field canting angle for which $R_{2\omega}$ reaches its maximum value in the *zy* scan, for a current direction parallel to ***k***. This provides a new way, alternative to SR-ARPES, to probe the 3D spin texture of the TSS via transport measurements.

**Transport measurements of the helical spin texture with hexagonal warping**

In Fig. 3a-c, we note that in the *xy* plane $R_{2\omega}$ retains the same angular dependence regardless of the orientation of the current along three typical directions in ***k***-space, namely $\overline{\Gamma K}$, $\overline{\Gamma M}$ and $\overline{\Gamma K'}$. This is consistent with the rotational symmetry of the *xy* plane projection of the surface spin texture. $R_{2\omega}$ approaches zero when the magnetic field is aligned with the direction of the current (i.e., $\varphi = 0°$ or $180°$), has a peak when the current and the magnetic field are orthogonal (i.e., $\varphi = 90°$ or $270°$), and changes sign when the magnetic field is reversed. These observations are consistent with the fact that the spin *s* orientation on the Fermi contour of the TSS is always perpendicular to the corresponding wave vector $k$[7-16], i.e., $s(k) \perp k$, and that the spin directions are opposite for opposite wave vectors, i.e., $s(-k) = -s(k)$, due to the spin-momentum locking.

More remarkably, we observe that the out-of-plane field scans of $R_{2\omega}$, in contrast to the in-plane ones, exhibit a striking dependence on the direction of the current. For the *zx* plane scans, a sizable angular dependence is found when the current is along the $\overline{\Gamma K}$ and $\overline{\Gamma K'}$ line in Figs. 3a and 3c, with peaks occurring when the magnetic field is perpendicular to the surface (i.e., $\theta = 0°$ or $180°$). Moreover, $R_{2\omega}$ switches the sign when the current direction is rotated from $\overline{\Gamma K}$ to $\overline{\Gamma K'}$. When the current direction is along the $\overline{\Gamma M}$ line (forming an angle of 30° with the $\overline{\Gamma K}$ line, see Fig. 3b), $R_{2\omega}$ becomes negligible in the *zx* scan. These observations are in qualitative agreement with a hexagonally warped helical spin texture, in which the out-of-plane component of the spin changes sign when $k$ is rotated from $\overline{\Gamma K}$ to $\overline{\Gamma K'}$ and vanishes when $k$ is along $\overline{\Gamma M}$ [10, 12-16] (see Fig. 1b). In the presence of hexagonal warping, the full rotational symmetry of the spin texture on the Fermi contour around the *z*-axis is reduced to a lower $C_{3v}$ symmetry, as first predicted theoretically[10] and later observed by SR-ARPES[12-16]. Thus, our $R_{2\omega}$ results indicate the existence of hexagonal warping on the Fermi surface.

For a quantitative comparison, we compare the magnetic field canting angle $\psi_{Hc}$ with the spin canting angle $\psi_{sc}$ in the TSS. We define $\psi_{Hc}$ as the angle between the *xy* plane and the direction of the magnetic field for which $R_{2\omega}$ reaches its maximum value in the $zy$ scan for a given current direction. $\psi_{sc}$ is defined as the angle between the $xy$ plane and the orientation of the spin in the TSS with $k$ parallel to the current. The reason for selecting field scans in the $zy$ plane is that the spin of the TSS with $k$ parallel to the current (which is chosen as the $x$-axis) always lies in the $zy$ plane due to spin-momentum locking. The field canting angle is obtained from $\psi_{Hc} = 90° - \theta_{pp}$, where $\theta_{pp}$ stands for the field angle corresponding to the positive peak position of $R_{2\omega}^{yz}$, which is quite different for the three current directions in Figs. 3a-c. In Fig. 4a, we show the field canting angle $\psi_{Hc}$ as a function of the angle between the direction of the current and the $\overline{\Gamma K}$ line (denoted by $\phi_{\Gamma K}$). The angular dependence is sinusoidal and has a period of 120°, tallying with the $C_{3v}$ symmetry of the spin texture in $Bi_2Se_3$ with hexagonal warping[13, 16].

On the other hand, the out-of-plane spin polarization derived from $k \cdot p$ perturbation theory is[10]

$$s_z(\phi_k) = \frac{\cos 3\phi_k}{\sqrt{(\cos 3\phi_k)^2 + (\alpha \hbar / \lambda k_F^2)^2}} \quad (1)$$

where $\phi_k$ is the angle between the $\overline{\Gamma K}$ line and the $k$ direction that coincides with the current direction, $k_F$ is the magnitude of the Fermi wave vector, $\alpha$ is the Dirac velocity, and $\lambda$ characterizes the strength of the hexagonal warping[10-16]. The $k$-dependent spin canting angle can be calculated by $\psi_{sc}(\phi_k) = \arcsin(s_z)$. In Fig. 4b, we show the $\psi_{sc}$ as a function of $\phi_k$ as a best fit to the experimental data given in Fig. 4a, obtained by using the following material parameters: $\alpha = 5 \times 10^5$ m/s, $\lambda = 180$ eV·Å$^3$ and the Fermi energy $\varepsilon_F = 0.28$ eV ($k_F \cong \varepsilon_F / \alpha \hbar$), which are in quantitative agreement with the values obtained by recent measurements[11, 13, 16]. Thus, we have demonstrated a direct correlation between the angular dependence of $R_{2\omega}$ in the TI and the 3D spin texture of the TSS with hexagonal warping.

**Physical origin of bilinear magneto-electric resistance**

Nonlinear UMR has been recently observed in heavy metal/ferromagnet or TI/magnetic-TI bilayers[30-34]. The presence of a ferromagnetic layer has been deemed to play an essential role as a source of spin-dependent scattering[30-34]. In our system, however, the nonlinear $R_{2\omega}$ occurs in a nonmagnetic $Bi_2Se_3$ single layer, with no ferromagnetic material involved and no reliance on spin-dependent scattering. In order to further investigate the physical origin of the observed magneto-electric resistance, we have measured $R_{2\omega}$ as a function of the amplitudes of the current $I$ (proportional to the applied electric field) and the magnetic field $H$. As shown in Fig. 5a, the amplitude $\Delta R_{2\omega}$ of the angular dependent $R_{2\omega}$ exhibits a linear dependence on the current amplitude up to 0.75 mA. This feature is shared by the UMR in the bilayers with magnetic materials. In Fig. 5b, $\Delta R_{2\omega}$ increases linearly with the amplitude of the magnetic field $H$ up to 9 T and displays a negligible value at zero magnetic field. The linear $H$ dependence of $R_{2\omega}$ is different from that of UMR, which is constant as a function of $H$ for a saturated FM layer[31-33].

The linear dependences of $R_{2\omega}$ on *both* the current (electric field) and the magnetic field is a new effect reported in TIs, which we refer to as the bilinear magneto-electric resistance. To further prove that the bilinear magneto-electric resistance is linked to spin-momentum locked TSS with hexagonal warping, we have theoretically calculated the longitudinal resistivity by solving the homogeneous Boltzmann equation with the model

Hamiltonian for an electron in a spin-momentum locked band of the form $H_{TI} = \boldsymbol{\sigma} \cdot [\boldsymbol{h}(\boldsymbol{k}) + g\mu_B \boldsymbol{H}]$, with the momentum ($\boldsymbol{k}$)-dependent field given by

$$\boldsymbol{h}(\boldsymbol{k}) = \alpha \hbar \boldsymbol{k} \times \hat{\boldsymbol{z}} + \lambda \boldsymbol{k} \times \hat{\boldsymbol{y}}'(k_x^2 - 3k_{y'}^2), \qquad (2)$$

where $\hat{\boldsymbol{y}}'$ and $\hat{\boldsymbol{z}}$ are the unit vectors denoting the directions along $\overline{\Gamma M}$ and normal to the surface respectively, $\boldsymbol{\sigma}$ are the Pauli spin matrices, and the term cubic in momentum describes the hexagonal warping effect[10]. We obtain an analytical expression of $R_{2\omega}$ by expanding the solution of the Boltzmann equation to second order in the electric field and to first order in the magnetic field (see Supplementary Information S6 for detailed derivation). The result is

$$R_{2\omega} = E \frac{l}{w}(a_{IP} H_y + a_{OP} H_z \cos 3\phi_{\Gamma K}), \qquad (3)$$

where $H_y$ and $H_z$ are the $y$ and $z$ components of the magnetic field defined in Fig. 2a-c, and $\phi_{\Gamma K}$ is the angle of a given $\overline{\Gamma K}$ direction with respect to the $x$-axis (i.e., the direction of the electric field $\boldsymbol{E}$), $a_{IP} = \dfrac{36\pi\lambda^2 \varepsilon_F g\mu_B}{e\alpha^5 \hbar^6}$ and $a_{OP} = \dfrac{6\pi\lambda g\mu_B}{e\alpha^2 \hbar^3 \varepsilon_F}$ are the coefficients of the contributions to $R_{2\omega}$ from the in-plane and out-of-plane components of the magnetic field, respectively. The theoretical formula for $R_{2\omega}$ shows a linear dependence on both the electric field and magnetic field, in line with our experimental findings. In Fig. 3d-f, theoretical fits to the experimental data are shown obtained by using Eq. (3). The excellent agreement between theory and experiment confirms that the spin-momentum locked TSS with hexagonal warping is indeed shaping the magneto-electric resistance.

Microscopically, the bilinear magneto-electric resistance can be interpreted as a partial conversion of a pure spin current of second order in the applied electric field $\boldsymbol{E}$ into a charge current in the TSS. In the absence of the magnetic field, the electron group velocity $v$ is odd in $k$ whereas the second order electron distribution function $\delta f_2(k)$ is even in $k$, i.e., $v(-k) = -v(k)$ and $\delta f_2(-k) = \delta f_2(k)$ (see Fig.1c). On the other hand, electrons at $k$ and $-k$ carry opposite spins due to spin-momentum locking. As a result, in the absence of a magnetic field, there are equal numbers of electrons carrying opposite spins and moving in opposite directions, which gives rise to a pure spin current $J_s(E^2)$ at the second order in the electric field $E$, as schematically shown in Fig. 1d. When a magnetic field is applied, however, both the group velocity and the second order distribution are shifted in $k$-space and then the two fluxes of electrons with opposite spin orientations no longer compensate each other, causing the spin current to be partially converted into a charge current $J'_e(E^2)$, as shown in Fig. 1e,f. This extra charge current, which changes with the relative orientation of the magnetic field with respective the electric field, is the physical reason for the bilinear magneto-electric resistance.

Finally, we note that the spin-charge current conversion will *not* take place when the band energy is linear in $k$, since in this case, the magnetic field changes both the electron velocity and the second order distribution through a constant displacement in momentum space: the two changes cancel against each other leaving the current unaltered.

**Summary and outlook**

We have shown that measurements of a magneto-electric resistance that depends linearly on both the electric and the magnetic field provide a simple way to probe the 3D spin texture of TSS with hexagonal warping by purely electrical means and offers an alternative to SR-ARPES. Our experimental observations, together with the theoretical calculations, reveal the critical role of spin current to charge current conversion occurring at the TSS under a magnetic field in shaping the bilinear magneto-electric resistance. Our method can also be applied to materials with giant Rashba spin splitting at the surface or in the bulk[35], as well as two-dimensional transition metal dichalcogenides with spin-polarized valley states[36]. The magneto-electric resistance is an effect of potential technological interest, similar to the UMR recently observed in magnetic heterostructures, but significantly different from it for being linear in the magnetic field and not dependent on the presence of ferromagnetic elements and spin-polarization-dependent scattering rates.

**Methods**

**Sample preparation.** The high-quality 20 QL $Bi_2Se_3$ films were grown on $Al_2O_3$ (0001) substrates in a molecular beam epitaxy system with a base pressure $< 2\times10^{-9}$ mbar using the two-step deposition procedure[37]. Before sample growth, the sapphire substrates were annealed at 750 °C for 30 min in a vacuum after transferring into the growth chamber. High-purity Bi (99.9999%) and Se (99.999%) were evaporated from standard Knudsen cells using Se/Bi flux ratio of ~20. To reduce Se vacancies in $Bi_2Se_3$, initial 2-3 QL $Bi_2Se_3$ were first deposited at 150 °C, and then the substrate temperature was ramped to 250 °C at 5 °C/min under Se flux for the second step growth. For the transport measurements, a capping layer of MgO (2 nm)/$Al_2O_3$ (3 nm) was deposited on the $Bi_2Se_3$ films as a protection layer. Hall bar devices with channel dimensions of $20\times100$ µm$^2$ were fabricated using the standard photolithography method.

**Electrical measurements.** Hall bar devices were wire-bonded to the sample holder and installed in a physical property measurement system (PPMS, Quantum Design) for transport measurements with the temperature range of 2-400 K. We performed the measurements of a.c harmonic longitudinal resistance, using a Keithley 6221 current source and two Stanford Research SR830 lock-in amplifiers. During the measurements, a constant amplitude sinusoidal current with a frequency of 251 Hz is applied to the devices, and the in-phase first harmonic $V_\omega$ and out-of-phase second harmonic $V_{2\omega}$ longitudinal voltage signals were measured simultaneously by two lock-in amplifiers, while rotating the magnetic field ***H*** in the *xy*, *zy* and *zx* planes. For a.c longitudinal resistance measurements under a dc bias current, Keithley 6221 provided a dc current with a certain polarity and an a.c. current with a small amplitude (50 µA) simultaneously, and Stanford Research SR830 lock-in amplifier was used to measure the in-phase first harmonic voltage $V_\omega$. For dc measurements of the longitudinal resistance, Keithley 2400 and 2002 were used.

**Acknowledgments**

This work was partially supported by the A*STAR's Pharos Programme on Topological Insulators, Ministry of Education-Singapore Academic Research Fund Tier 1 (R-263-000-B47-112) and NSF Grant DMR-1406568.

**Author contributions**

P.H. and H.Y. planned the study. D.Z. fabricated devices. P.H. and D.Z. measured transport properties. Y.L., Y.W. and J.Y. helped characterization. S.S.-L.Z and G.V did the theory. All authors discussed the results. P.H., S.S.-L.Z, D.Z., G.V and H.Y. wrote the manuscript. H.Y. supervised the project.

**Additional information**

Supplementary information is available in the online version of the paper. Reprints and permissions information is available online at www.nature.com/reprints. Correspondence and requests for materials should be addressed to H.Y.

**Competing financial interests**

The authors declare no competing financial interests.

**Figure captions**

**Figure 1| Schematic of the spin texture of surface states of a 3D topological insulator with hexagonal warping and conversion of spin current to charge current by applying an external magnetic field. a**, Hexagonally warped energy dispersion for the surface states with Fermi surface lying in the conduction band. **b**, Hexagonally warped spin texture at the Fermi contour of the surface states. **c,** Schematics of the variation of the electron distribution along the **k**-axis parallel to the applied electric field **E**: δ$f_1$ (blue curve) and δ$f_2$ (yellow curve) are the corrections to the equilibrium distribution at the first and second order of the electric field, respectively; the solid arrows represent excess of electrons with their spins along the arrow direction, while the hollow arrows represent depletion of the same. **d**, When an electric field **E** is applied along a certain direction in **k**-space (dash-dotted line), a non-equilibrium spin current $\boldsymbol{J}_s(\boldsymbol{E}^2)$ is generated at the second order of the electric field, due to spin momentum locking. **e-f,** When an external magnetic field is applied, the nonlinear spin current is partially converted into a charge current $\boldsymbol{J}'_e(\boldsymbol{E}^2)$: a high resistance state is reached when the magnetic field is parallel (**e**) to the spin direction of the spin current, while a low resistance state is reached when it is antiparallel (**f**).

**Figure 2| First ($R_\omega$) and second ($R_{2\omega}$) harmonic magneto-resistances in three different geometries.** Geometries of longitudinal magneto-resistance measurements with rotating *H* in the *xy* (**a**), *zy* (**b**) and *zx* (**c**) planes. Typical first (blue) and second (red) harmonic resistances measured for a 20 QL Bi$_2$Se$_3$ device under *H* = 9 T, *T* = 125 K and *I* = 0.55 mA along the $\overline{\Gamma K}'$ direction. The solid lines are fits to the data. A vertical offset in $R_{2\omega}$ was subtracted for clarity.

**Figure 3| Detection of the hexagonally warped helical spin texture of TSS. a-c**, Second harmonic resistance for all three scans for the devices with the current applied at an angle of 0° (**a**), 30° (**b**) and 60° (**c**) with respect to the $\overline{\Gamma K}$ direction, respectively. These scans were performed under $H = 9$ T, $T = 60$ K and $I = 0.55$ mA. Blue hexagons in **a-c** represent the Fermi surface of $Bi_2Se_3$. The red lines are along the $\overline{\Gamma K}$ direction and the black arrow denotes the current direction in *k*-space. **d-f**, The calculated angular dependences of $R_{2\omega}$ corresponding to the three cases in **a-c**, respectively. The parameters used in the calculation are $\alpha = 5 \times 10^5$ m/s, $\lambda = 165$ eV·Å$^3$, $\varepsilon_F = 0.256$ eV, $g = 2$, $H = 9$ T, $E = 100$ V/cm, $l = 100$ μm and $w = 20$ μm.

**Figure 4| Quantitative comparison between the field and spin canting angles. a**, The measured field canting angle as a function of the angle between the current direction and the $\overline{\Gamma K}$ line. **b**, The theoretical spin canting angle for a *k*-direction that coincides with the direction of the current.

**Figure 5| Current and magnetic field dependence of $R_{2\omega}$. a,b**, Current (**a**) and magnetic field (**b**) dependence of the amplitude $\Delta R_{2\omega}$ in the *zy* scan at 60 K for a device with the current along $\overline{\Gamma K'}$. The solid black lines are fits to the data.

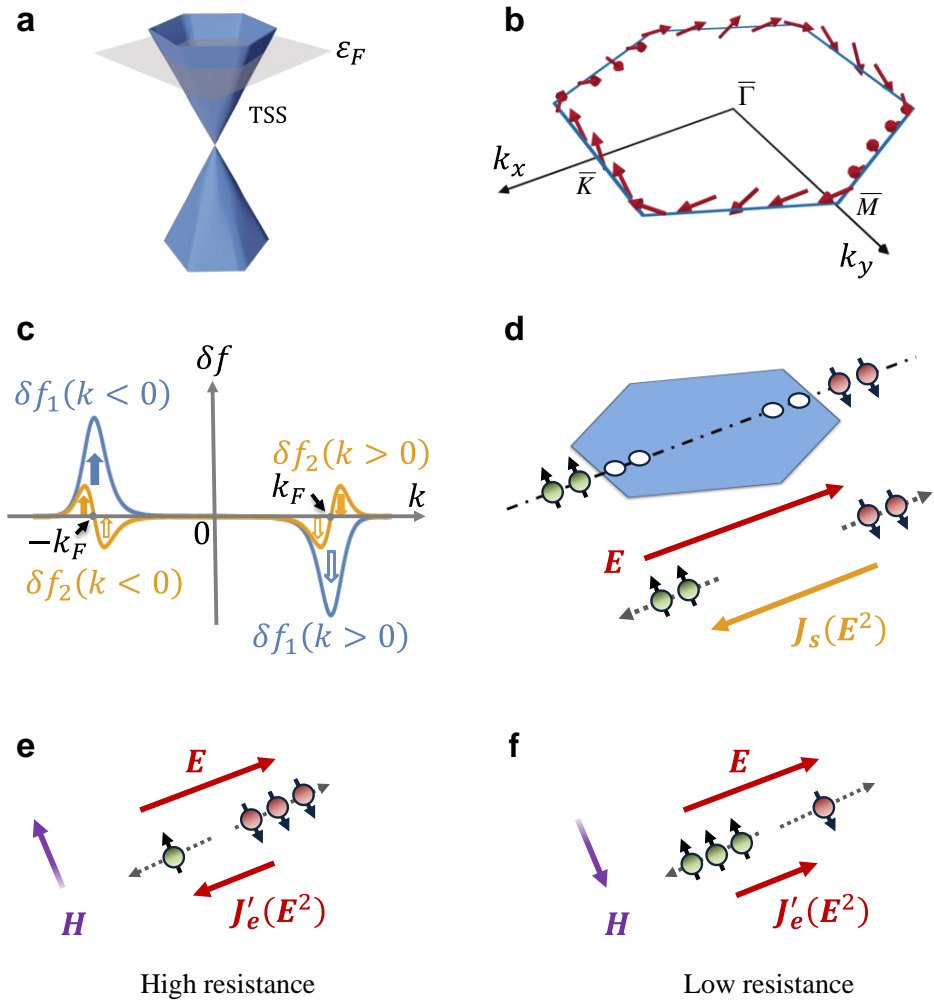

Figure 1

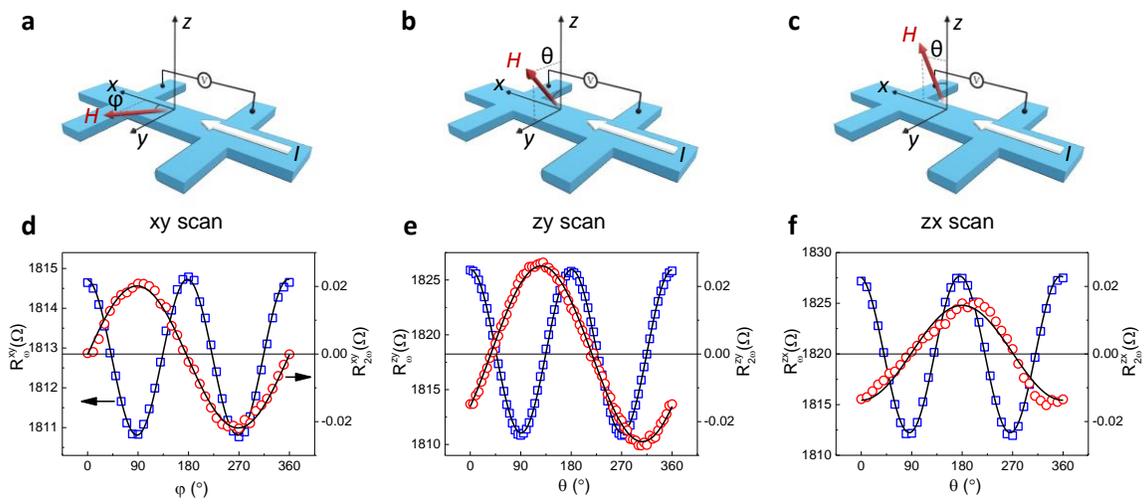

Figure 2

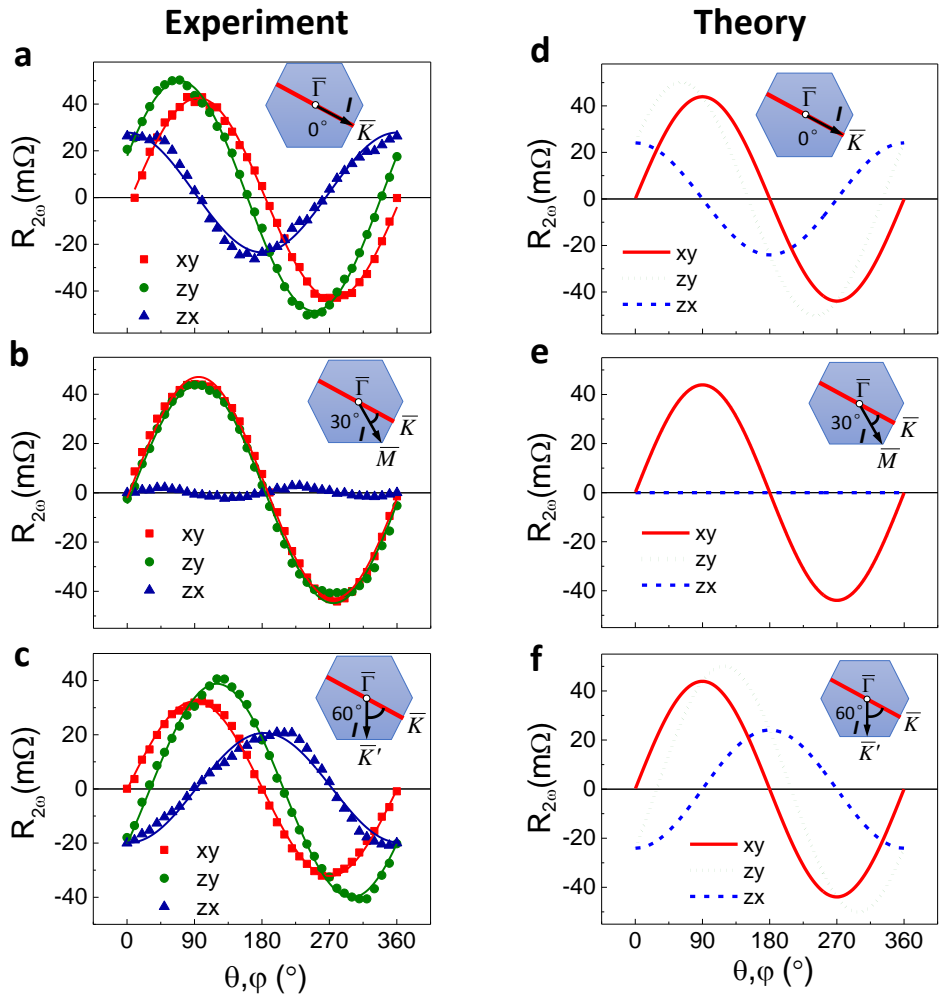

Figure 3

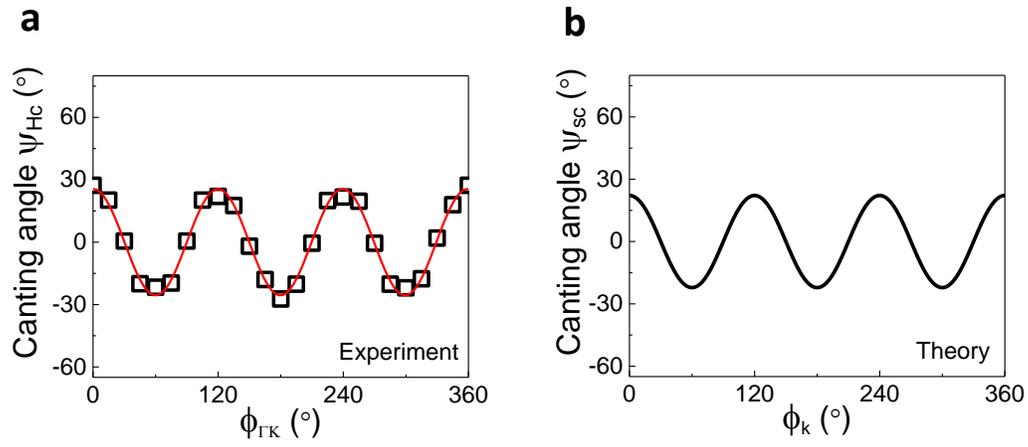

Figure 4

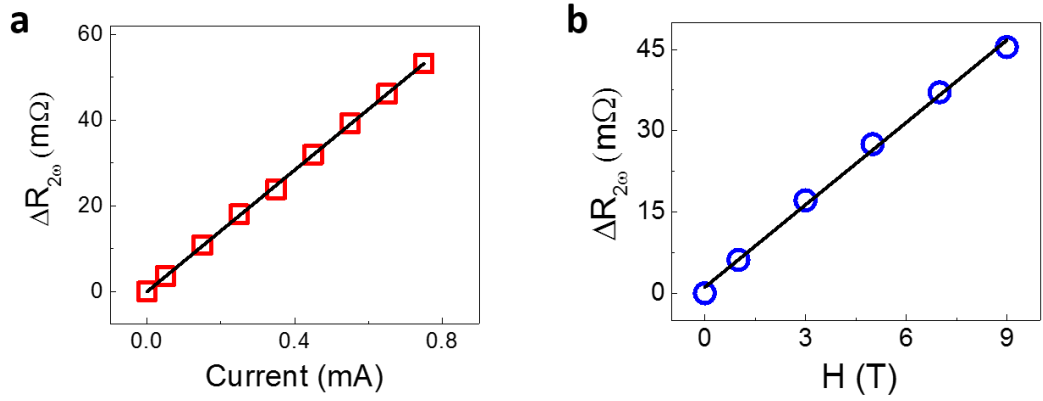

Figure 5